\documentclass[12pt]{article}

\usepackage{amsmath,amsthm,amsfonts,helvet}

\usepackage{cite}

\theoremstyle{plain}

\newtheorem{prop}{Proposition}
\newtheorem{lemma}[prop]{Lemma}

\newtheorem{theorem}[prop]{Theorem}

\theoremstyle{remark}
\newtheorem{rem}[prop]{\bf Remark}
\newtheorem{example}[prop]{\bf Example}

\numberwithin{equation}{section}

\newcommand{\spec}[1]{\mathop{\mathrm{spec}}(#1)}
\newcommand{\res}[1]{\mathop{\mathrm{res}}(#1)}

\DeclareMathOperator{\supess}{sup\,\,ess}
\DeclareMathOperator{\supp}{supp}

\newcommand{\RR}{\mathbb{R}}
\newcommand{\CC}{\mathbb{C}}

\newcommand{\NN}{\mathbb{N}}
\newcommand{\SSS}{\mathbb{S}}
\newcommand{\KK}{{\mathcal{K}\,}}

\newcommand{\dint}{\displaystyle\int}

\begin{document}

{
\raggedright

{\LARGE\bf\sffamily
On-diagonal singularities of the Green functions for~Schr\"odinger
operators

}

\vskip 7mm

\leftskip 20mm

{\large\sf
Jochen Br\"uning$^1$, Vladimir Geyler$^2$, Konstantin Pankrashkin$^{1,\,*}$}

\vskip 7mm

\leftskip 35mm

{
\it

$^1$ Institut f\"ur Mathematik, Humboldt-Universit\"at zu Berlin,
Rudower Chaussee 25, 12489 Berlin, Germany

$^2$ Mathematical Faculty, Mordovian State University, 430000 Saransk, Russia

$^*$ Corresponding author; e-mail:} \texttt{\small const@mathematik.hu-berlin.de}

}

\vskip 7mm

\leftskip 20mm

\noindent\textbf{Abstract.} \small We investigate the behavior of the Green functions of Schr\"odinger operators
near the diagonal. The only non-trivial cases, where the
on-diagonal singularities
are non-zero and do not depend on the spectral parameter,
are two and three dimensions.
In the case of two dimensions, we show that the singularity is independent
of both the scalar and the gauge potentials. In dimension three,
we obtain conditions for preserving the singularity under perturbations
by non-regular potentials. Some examples illustrating dependence
of the singularity on general scalar and gauge potentials are presented.

\leftskip 0mm

\bigskip

\section{Introduction}

Singularities of the Green functions of the quantum-mechanical
operators play a crucial role in many branches of theoretical and
mathematical physics, from which one should mention first the
renormalization procedure of the quantum field theory \cite{Wei,BS,Jac}. From the point of view of the high-derivative quantum
gravity, the corresponding problem was considered e.g. in
\cite{Avr}. In particular, in the case of non-minimal coupling of
quantum matter to the gravitational background with conical
singularities, an operator of the form $H=-\Delta+U$ arises on a
Riemannian manifold $X$. Here $\Delta$ is the Laplace-Beltrami
operator on $X$ and $U$ represents the non-minimal coupling term
$\xi \mathcal{R}$ with the Ricci scalar $\mathcal{R}$. The scalar
curvature possesses a distributional behavior at conical
singularities \cite{Sol}, $\mathcal{R} = \mathcal{R}_{\rm
reg}+4\pi(1-\alpha)\delta_M\,$, where $\delta_M$ is a Dirac
$\delta$-like potential supported by a sub-manifold $M\subset X$
and $2\pi(1-\alpha)$ is the angle deficit. As a result, an
operator
\begin{equation}
             \label{delta}
H_M=-\Delta+V+a\delta_M
\end{equation}
arises with the coupling constant $a=4\pi(1-\alpha)\xi$
characterizing the interaction with a background field
concentrated on $M$. Operators of such form appear in the
investigation of scalar fields with non-minimal coupling on the
cosmic string background, in the Euclidean approach to the black
hole thermodynamics, in the study of the particle scattering at
the Planck scale (see \cite{Sol} and references therein).
Moreover, in the context of the scattering theory, the potential
$V$ can have singularity (e.g. of the Coulomb type) even in the
case of a flat manifold $X$.

We are interesting here in the singular term $\delta_M$
concentrated on a zero-dimensional submanifold $M$; this case
covers not only quantum fields with point interactions, but also
the case when $M$ has a cartesian complement in $X$: $X=Y\times
M$. If $M$ is a uniformly discrete subset of $X$ and $\dim X\le3$,
then the Green function $G_M(x,y;\zeta)$ of $H$ can be obtained
through the Krein resolvent formula in terms of the Green function
$G(x,y;\zeta)$ for the operator $H=-\Delta+V$ \cite{Pav}. An
important ingredient of this formula is the so-called ``Krein
$\mathcal{Q}$-matrix'' (a kind of the Dirichlet-to-Neumann map)
$Q_{mn}(z)$, $m,n\in M$. To define the diagonal elements of $Q$
for $\dim X>1$, a renormalization procedure is needed. For smooth
$V$, the renormalized Green function $G^\mathrm{ren}(x,y;\zeta)$,
which must be continuous in the whole $X\times X$, is defined as
\begin{equation}
                             \label{stsing}
G^\mathrm{ren}(x,y;\zeta)=G(x,y;\zeta)-S(x,y)\,,
\end{equation}
where the ``standard singularity'' $S$ has the form $\displaystyle
S(x,y)=-\frac{1}{2\pi}\log d(x,y)$ if $\dim X=2$, and
$\displaystyle S(x,y)=\frac{1}{4\pi d(x,y)}$ if $\dim X=3$ (here
$d(x,y)$ is the geodesic distance on $X$). Now one can put
$Q_{mm}(\zeta)=G^\mathrm{ren}(m,m;\zeta)$. The corresponding
renormalization procedure in the Euclidean case is known long ago,
see e.g. \cite{DO} and \cite{AGHH} for the history and the quantum
mechanical treatment. It is important to note that usually one
obtains $S(x,y)$ by a momentum cutoff (an ultraviolet
regularization procedure); the result is equivalent to that
obtained with the help of a dimensional regularization. In the
case of brane coupling to gravity or to a gauge field it is
necessary to use a dimensional regularization \cite{GW}. It is
worthy to add that the strict mathematical treatment of the
operators (\ref{delta}) has its origins in the article \cite{BF}
by F.~Berezin and L.~Faddeev.

In the case ${\rm dim}\,X\ge4$ there is no regularization
procedure involving a singularity independent of the energy
parameter $\zeta$ (see Example~\ref{4d} below). Moreover, if $V$
has a Coulomb-like singularity or if an interaction with a gauge
field is present, then the function $S$ in (\ref{stsing}) is
different from the standard one, i.e., $\displaystyle S(x,y)\ne
\frac{1}{4\pi d(x,y)}$ (see Examples \ref{ex-c} and~\ref{ex-3d}
below); similar phenomena related to propagation of waves in
strongly inhomogeneous media have been studied recently
in~\cite{haba}. The main goal of our paper is to investigate the
situation in detail. We show that in dimension two the singularity
of $G$ has the standard form even in the presence of an additional
$\mathbf{U}(1)$-gauge potential (Theorem~\ref{main-th-2d}). On the
other hand, in dimension three, $S$ depends on $V$ modulo a
Lebesgue class of functions on $X$ (see Theorem~\ref{main-th-3d})
and is defined up only to a continuous additive term (the
situation here is completely similar to that for the Krein
$\mathcal{Q}$-functions: they are defined up to an additive
constant). The concrete value of this term is subject of analysis
of a given physical problem and is out of the scope of the present
work. We mention only that a possible way to fix the corresponding
additive constant is to compare the integrated density of states
with the trace of $G^\mathrm{ren}$. It is worthy to note that the
Green function for operators of the form (\ref{delta}) on a
three-dimensional Riemannian manifold was used recently for
simulating the confinement potential of a quantum dot in
\cite{GK}. Earlier the defects in solids are investigated by
methods of quantum gravity in \cite{KV}. New technologies of
manufacturing two-dimensional nanostructures with non-trivial
geometry \cite{TTOIYH,PGBDKD} caused the appearance of
mathematical models of such structures where, in particular, the
Hamiltonian has the form (\ref{delta}) with the $\delta$-term
simulating the potential of a short range impurity \cite{AGM}. If
the nanostructure is displaced in a magnetic field we must replace
$\Delta$ in (\ref{delta}) by the Bochner Laplacian. In this case
the properties of the Green function $G$ are needed for
investigation of explicitly solvable models of the geometric
scattering theory \cite{BG} or spectral theory of periodic hybrid
manifolds \cite{BEG}.

At last but not at least we stress that our main results are new
\emph{even for the case of Euclidean spaces $X=\RR^n$}.

\section{Definitions and preliminaries}

Throughout the paper we denote by $X$ a complete connected
Riemannian manifold of bounded geometry, which means that the
injectivity radius $r_\mathrm{inj}$ of $X$ is positive and every
covariant derivative of the Riemann curvature tensor is bounded.
Examples are provided by homogeneous spaces with invariant metrics
(in particular, Euclidean spaces), compact Riemannian manifolds
and their covering manifolds;  for discussion of various
properties of such manifolds in the context of differential
operators we refer to~\cite{Shub}. The dimension of $X$ we denote
by $\nu$; the geodesic distance between $x,y\in X$ will be denoted
by $d(x,y)$. For $x\in X$ and $r\ge 0$ we use the notation
$B(x,r)=\{y\in X:\, d(x,y)<r \}$; through the paper, we suppose
$r<r_{\rm inj}$ for radiuses $r$ of all considered sufficiently
small balls. For a measurable function $f$ on $X$, we denote by
$\|f\|_p$ the $L^p$-norm of $f$. If $K$ is a bounded operator from
$L^p(X)$ to $L^q(X)$, $1\le p,q\le\infty$, then its norm will be
denoted by $\|K\|_{p,q}$.

Let $A=A_j\,dx^j$ be a 1-form on $X$, for simplicity we suppose
here $A_j\in {C}^\infty(X)$. The functions $A_j$ can be considered
as the components of the vector potential of a magnetic field on
$X$. On the other hand, $A$ defines a connection $\nabla_A$ in the
trivial line bundle $X\times \mathbb{C}\rightarrow X$,
$\nabla_Au=du+iuA$; by $-\Delta_A=\nabla_A^*\nabla_A$ we denote
the corresponding Bochner Laplacian. In addition, we consider a
real-valued scalar potential $U$ of an electric field on $X$. This
potential will be assumed to satisfy the following conditions:
\begin{gather*}
U_+:=\max(U,0)\in {L}^{p_0}_\text{loc}(X),
\quad U_-:=\max(-U,0)\in \sum_{i=1}^n {L}^{p\,{}_i}(X),\\
\quad 2\le p_i\le\infty \text{ if }\nu\le3,
\quad \nu/2<p_i\le \infty \text{ if }\nu\ge4,\quad 0\le i\le n;
\end{gather*}
we stress that $p_i$ as well as $n$ are not fixed and depend on
$U$. The class of such potentials will be denoted by
$\mathcal{P}(X)$. Below we will need an approximation of singular
potentials by smooth ones; for this purpose the following lemma is
useful.

\begin{lemma}
                           \label{lmain}
Let $f\in L^p_{\rm loc}(X)$, where $1\le p<\infty$, and $f\ge 0$.
Then there is $g\in C^\infty(X)$ such that $g\ge 0$ and
$\displaystyle f-g\in L^q(X)$ for all $1\le q\le p$.
\end{lemma}

\begin{proof}
Fix $a\in X$ and for integers $n$, $n\ge1$, denote
$Y_n=B(a,n)\setminus\bar B(a,n-1)$. Fix a real sequence $a_n$,
$a_n>0$ such that $\sum a_n\le1$ and denote by $f_n$ the
restriction of $f$ to the set $Y_n$. Since the measure of $Y_n$ is
finite, for every $n$ we can find a function $g_n$, $g_n\in
C_0^\infty(X)$, such that $g_n\ge0$, ${\rm supp}\,(g_n)\subset
Y_n$, and $\max(\|f_n-g_n\|^p_p,\,\|f_n-g_n\|_1)\,\le a_n$. Since
the family $(Y_n)$ is locally finite, the point-wise sum $g=\sum
g_n$ exists and $g\in C_0^\infty(X)$. It is clear that $g\ge0$ and
$\max(\|f-g\|_p\,,\|f-g\|_1)\,\le1$, i.e., $f-g\in L^p(X)\cap
L^1(X)$. \qed\end{proof}

We denote by $H_{A,U}$ the operator acting on functions $\phi\in
{C}_0^\infty(X)$ by the rule $H_{A,U}\phi=-\Delta_A\phi+U\phi$.
This operator is essentially self-adjoint in ${L}^2(X)$ and
semibounded below~\cite{BGP}; its closure will be also denoted by
$H_{A,U}$. By $\spec{H_{A,U}}$ we denote the spectrum of $H_{A,U}$
and by $\res{H_{A,U}}$ the set of regular points:
$\res{H_{A,U}}=\CC\setminus\spec{H_{A,U}}$. Let us denote the
resolvent of $H_{A,U}$ by $R_{A,U}(\zeta)$, i.e.
$R_{A,U}(\zeta)=(H_{A,U}-\zeta)^{-1}$.

Here we introduce two classes of integral kernels used in the
paper. First class, $\KK_\text{cont}(p)$, $1\le p\le\infty$,
consists of all continuous on $X\times X$ functions $K(x,y)$
satisfying for any $r>0$ the condition
\begin{equation}
         \label{L2}
\lfloor K\rfloor_{p,r}:=\max\Big(\,\supess\limits_{x\in X}
\big\|\chi_{X\setminus B(x,r)}\,K(x,\cdot)\big\|_p\,,
\supess\limits_{y\in X} \big\|\chi_{X\setminus
B(y,r)}\,K(\cdot,y)\big\|_p\,\Big) <\infty\,,
\end{equation}
where $\chi_A$ stands for the characteristic function of the set
$A\subset X$. The second class, $\KK(\alpha,p)$, $0\le\alpha<\nu$,
$1\le p\le\infty$, consists of all measurable functions $K$ on
$X\times X$ obeying the condition~\eqref{L2} and
\begin{equation}
           \label{L1}
\big|K(x,y)\big|\le c \max(1,d(x,y)^{-\alpha}) \text{ for a constant } c=c(K)>0.
\end{equation}
We put $\KK_\text{cont}(\alpha,p):=\KK(\alpha,p)\cap
{C}(X\times X\setminus D)$, where $D$ is the diagonal $\big\{(x,y)\in X\times X:\,x=y\big\}$.

The above introduced classes of integral kernels are important due
to their relations to the properties of the resolvents
$R_{A,U}(\zeta)$; these relationships are stated in the following
theorem which is our starting point (see \cite{BGP} for the proof).

\begin{theorem}\label{th-bgp}
For any $\zeta\in\res{H_{A,U}}$ the resolvent $R_{A,U}(\zeta)$ has
an integral kernel $G_{A,U}(x,y;\zeta)$, the Green function, which
belongs to $\KK_\mathrm{cont}(\lambda,q)$, where $q$, $1\le
q\le\infty$, is arbitrary, and $\lambda=\nu-2$ for $\nu>2$,
$\lambda\in (0,\nu)$ is arbitrary for $\nu=2$, $\lambda=0$ for
$\nu=1$; moreover, $G_{A,U}$ is continuous in $X\times X$ for
$\nu=1$.
\end{theorem}

We should point out that the Green function of a Schr\"odinger
operator can violate the conditions~\eqref{L2} and~\eqref{L1}, if
the potential $U$ is not from the class $\mathcal{P}(X)$. Even the
decay of the Green function for large distances between $x$ and
$y$ (the off-diagonal behavior) can be different from the
``standard'' exponential one coming from the comparison with the
Laplacian; a good example is delivered by the one-dimensional
inverse harmonic oscillator, whose Green function has only a
polynomial decay at infinity (see Appendix~\ref{app-c}).

Our further calculations will involve a couple of operations with integral kernels
introduced above; here we
collect some useful estimates which will be used very intensively.

The well-known Gelfand-Dunford-Pettis theorem claims that if $K$
is a bounded operator from ${L}^p(X)$ to ${L}^\infty(X)$ with some
$p$, $1\le p<\infty$, then it is an integral operator and its
kernel $K(x,y)$ satisfies the estimate
\begin{equation}
            \label{est-s}
\supess\limits_{x\in X}\|K(x,\cdot)\|_{q}<\infty,\quad \quad
q=(1-p^{-1})^{-1}.
\end{equation}
Conversely, if a kernel $K(x,y)$ satisfies~\eqref{est-s}, then it is an
integral kernel of a bounded operator from ${L}^p(X)$ to
${L}^\infty(X)$.

\begin{lemma}
           \label{lem-sws}
Let $K_j:{L}^{q_j}(X)\to{L}^\infty(X)$, $1\le q_j<\infty$,  be
bounded linear operators with integral kernels $K_j(x,y)$,
$j=1,2$, and $W\in {L}^{q_1}(X)$, then for a.e $(x,y)\in X\times
X$ the integral $J(x,y)=\dint_X K_1(x,z) W(z) K_2(z,y)dz$ exists
and $J(x,y)$ is an integral kernel of the operator $K_1WK_2$.
\end{lemma}

\begin{proof}
The operator $K_1 W K_2$ is bounded from ${L}^{q_2}(X)$ to
${L}^\infty(X)$, therefore, it is an integral operator. Let $f\in
{L}^{q_2}(X)\cap C(X)$ such that $f(x)>0$ for all $x\in X$. Then
there holds
\begin{equation}
          \label{eq-sws}
K_1 W K_2 f(x) = \int_X K_1(x,z) W(z) \int_X K_2(z,y) f(y)dy\,dz.
\end{equation}
{}From the other side, according to the estimates~\eqref{est-s} for
$K_1$ and $K_2$, there holds
\[
\dint_X \big|K_2(\cdot,y) f(y)\big|dy\in{L}^\infty(X), \quad
\Big|W(\cdot)\Big|\dint_X \Big|K_2(\cdot,y)
f(y)\Big|dy\in{L}^{q_1}(X),
\]
hence,
\[
\int_X \big|K_1(x,z)\big| \bigg(\big|W(z)\big|\dint_X
\big|K_2(z,y) f(y)\big|dy\bigg)dz<\infty\,.
\]
By the Fubini
\[
\int_X  \Big(\int_X |K_1(x,z) W(z)  K_2(z,y)| dz\Big)\,
f(y)dy<\infty\,,
\]
and since $f(x)>0$, the inner integral exists for a.e. $(x,y)\in
X\times X$.

Let now $f$ be an arbitrary function from ${L}^{q_2}(X)$.
Repeating the arguments above, we get
\begin{equation}
          \label{eq-sws2}
K_1 W K_2 f(x) = \int_X  \Big(\int_X K_1(x,z) W(z)  K_2(z,y)
dz\Big)\, f(y)dy\,
\end{equation}
for a.e $x\in X$. Therefore $J$ is an integral kernel for
$K_1 W K_2$.
\end{proof}

We will often use the estimate given by the lemma below
(cf.~\cite{BGP}):
\begin{lemma}\label{oft}
There exists $r_0>0$ such that for any $\alpha,r$ with $0<r<r_0$,
$0\le\alpha<\nu$, and $a,x\in X$ there holds
\begin{equation}
                \label{est-j1}
\int_{B(a,r)}\frac{dy}{d(x,y)^\alpha}\le c r^{\nu-\alpha}
\end{equation}
with some $c>0$ depending only on $\alpha$.
\end{lemma}

Our next auxiliary result is the following lemma.

\begin{lemma}\label{lem-ka}
Let $K\in\KK(\alpha,p)$, $1\le p<\infty$, $p\alpha<\nu$, and
$1/p+1/q=1$, then $K$ is an integral kernel of a bounded operator
from ${L}^q(X)$ to ${L}^\infty(X)$.
\end{lemma}

\begin{proof}
According to the Gelfand-Dunford-Pettis theorem we must prove
\[
\supess_{x\in X}\int_X \big|K(x,y)\big|^p dy<\infty.
\]
Fix $r$, $0<r<r_0$, and for $x\in X$ expand the integral into two
parts:
\[
\int_X \big|K(x,y)\big|^p dy= \int_{B(x,r)} \big|K(x,y)\big|^p dy
+\int_{X\setminus B(x,r) } \big|K(x,y)\big|^p dy\,.
\]
The first term is estimated by Lemma~\ref{oft}, and the second one
is majorated  by $\lfloor K \rfloor^p_{p,r}$.
\end{proof}

\begin{lemma}\label{main}
Let three measurable functions $K_1(x,y)$, $K_2(x,y)$ and $W(x)$
be given, where $x,y\in X$. Denote
$F(x,y,z):=K_1(x,z)W(z)K_2(z,y)$, and if the integral $\dint_X
F(x,y,z)\,dz$ exists, denote it by $J(x,y)$.

\textup{(A)} Let $K_j\in \KK_\mathrm{cont}(\alpha_j,p_j)$,
$j=1,2$, and $W\in{L}^p(X)$, such that $1/p_1+1/p_2+1/p=1$ and
$p>\nu/\big(\nu-\max(\alpha_1,\alpha_2)\big)$. Then
$F(x,y,\cdot)\in {L}^1(X)$ for $x\ne y$, hence $J$ is well
defined. Moreover, $J\in\KK_\mathrm{cont}(\alpha,\infty)$, where
$\alpha=\max\big(p'(\alpha_1+\alpha_2)-\nu,0\big)$ with
$1/p+1/p'=1$, if $p'(\alpha_1+\alpha_2)\ne\nu$, and $\alpha$ is an
arbitrary number from $(0,\nu)$ otherwise.

\textup{(B)} Let the conditions of the item~\textup{(A)} be
satisfied. Assume additionally that $\alpha_1+\alpha_2<\nu$ and
$W\in {L}^q_\mathrm{loc}(X)$ with $q>\nu/(\nu-\alpha_1-\alpha_2)$.
Then $F(x,y,\cdot)\in {L}^1(X)$ for any $x,y\in X$ and $J\in
{C}(X\times X)$.

\textup{(C)} Let $W\in L^p(X)$, and
$K_1\in\KK_\mathrm{cont}(p_1)$,
$K_2\in\KK_\mathrm{cont}(\alpha,p_2)$ or
$K_1\in\KK_\mathrm{cont}(\alpha,p_1)$,
$K_2\in\KK_\mathrm{cont}(p_2)$. Assume additionally that
$1/p+1/p_1+1/p_2=1$ and $p>\nu/(\nu-\alpha)$. Then
$F(x,y,\cdot)\in {L}^1(X)$ for any $x,y\in X$, and $J\in
{C}(X\times X)$.
\end{lemma}

\begin{proof}

The proof of the items (A) and (B) is given
in~\cite{BGP}.

(C) We give a proof for the case $K_1\in\KK_\mathrm{cont}(p_1)$
and $K_2\in\KK_\mathrm{cont}(\alpha,p_2)$; the second case can be
considered exactly in the same way.

Let $x,y\in X$; we show first that $F(x,y,\cdot)\in{L}^1(X)$. Let
$r>0$, then for $z\in B(y,r)$ we have
\begin{equation}
     \label{fxy-loc}
\big|F(x,y,z)\big|\le c k_1(x,y) W(z) d(y,z)^{-\alpha}, \quad
k_1(x,y):=\sup_{z\in B(y,r)} K_1(x,z)<\infty, \quad c>0,
\end{equation}
therefore, $F(x,y,\cdot)\in{L}^1\big(B(y,r)\big)$ due to the
H\"older inequality and our conditions on $p$. For $z\notin
B(y,r)$ due to the H\"older inequality we have the estimate
\[
\int_{X\setminus B(y,r)}\big|F(x,y,z)\big|dz\le
\bigg(\int_{X\setminus
B(y,r)}\big|K_1(x,z)\big|^{p_1}dz\bigg)^{1/p_1}  \lfloor
K_2\rfloor_{p_2,r}\|W\|_{p}\,,
\]
and
\[
\int_{X\setminus B(y,r)}\big|K_1(x,z)\big|^{p_1}dz \le
\int_{X}\big|K_1(x,z)\big|^{p_1}dz
=\int_{B(x,r)}\big|K_1(x,z)\big|^{p_1}dz+\int_{X\setminus
B(x,r)}\big|K_1(x,z)\big|^{p_1}dz\,,
\]
where the first term on the right-hand side is finite due to the
continuity of $K_1$, and the second one is estimated
by~\eqref{L2}. This proves the inclusion
$F(x,y,\cdot)\in{L}^1(X)$.

Now let $x_0,y_0\in X$, $0<r<R$, and $x\in B(x_0,r/2)$, $y\in
B(y_0,r/2)$, then
\begin{multline}
        \label{eq-jj}
\big|J(x,y)-J(x_0,y_0)\big|\le \int_{B(y_0,r)}
\big|F(x,y,z)\big|dz+\int_{B(y_0,r)}
\big|F(x_0,y_0,z)\big|dx\\
{}+\int_{X\setminus B(y_0,R)}\big|F(x,y,z)\big|dz
+\int_{X\setminus B(y_0,R)}\big|F(x_0,y_0,z)\big|dz\\
{}+\int_{B(y_0,R)\setminus B(y_0,r)}\Big|
F(x,y,z)-F(x_0,y_0,z)\Big|dz.
\end{multline}
Take $\epsilon>0$ and assume $r<r_0$. For $z\in B(y_0,r)$ we
estimate $F(x,y,z)$ as in~\eqref{fxy-loc}, then we get using
Lemma~\ref{oft}
\[
\int_{B(y_0,r)} \big|F(x,y,z)\big|dz\le c
\sup_{\begin{subarray}{l}x\in B(x_0,r),\\y\in
B(y_0,r)\end{subarray}} K_1(x,y) \|W\|_p \left[\,\int_{B(y_0,r)}
d(y,z)^{\frac{p\alpha}{1-p}}\,dz\,\right]^{\frac{p-1}{p}} \le C
r^{\nu-\alpha-\frac{1}{p}}=o(1)
\]
as $r\to 0$. On the other hand
\[
\dint_{X\setminus B(x_0,R)}\big|F(x,y,z)\big|dz\le \lfloor
K_1\rfloor_{p_1,r}\lfloor K_2\rfloor_{p_2,r} \|\chi_{X\setminus
B(x_0,R)}W\|_p=o(1)\quad {\rm as}\quad R\to\infty\,.
\]
Finally, we conclude that $r$ can be taken sufficiently small and
$R$ sufficiently large, such that the sum of the first four terms
on the right-hand side of~\eqref{eq-jj} is less than $\epsilon/2$.
Now it is sufficient to prove that at these fixed $r$ and $R$ the
function
\[
\dint_{B(y_0,R)\setminus B(y_0,r)} F(x,y,z)dz
\] is continuous
as $x\in B(x_0,r/2)$ and $y\in B(y_0,r/2)$. To do this, we note
that with some $C'>0$ the following estimate
$\big|F(x,y,z)\big|\le C' \big|W(z)\big|$ takes place for all
$x\in B(x_0,r/2)$, $y\in B(y_0,r/2)$, and $z\in B(y_0,R)\setminus
B(y_0,r)$. Since $W\in {L}^1\big(B(y_0,R)\setminus B(y_0,r)\big)$,
the requested continuity follows from the Lebesgue majorization
theorem.
\end{proof}

As it was mentioned in the introduction, we are going to present
the Green function in the form
\[
G_{A,U}(x,y;\zeta)=S_{A,U}(x,y)+G^\mathrm{ren}_{A,U}(x,y;\zeta),
\]
where the second term must be continuous in $X\times X$.
Such a representation is trivial in the one-dimensional
case: the Green function is continuous, and one can put $S_{A,U}\equiv 0$.
In dimensions $\nu\ge 4$ the problem makes no sense, as the following example
shows:
\begin{example}[\textbf{Four-dimensional Laplace operator}]\label{4d}
Consider the simplest case of the Laplacian in ${L}^2(\RR^4)$.
The Green function takes the form
\[
G(x,y;\zeta)=\frac{\sqrt{-\zeta}}{4\pi^2|x-y|}K_1\big(\sqrt{-\zeta}
\, |x-y|\big),
\]
where $K_1$ is the modified Bessel function of the first order. Near the
diagonal $x=y$ one has
\[
G(x,y;\zeta)=\frac{1}{4\pi^2|x-y|^2}-\frac{\zeta\log |x-y|}{8\pi^2}
+k(x,y;\zeta)
\]
with a continuous $k$. Therefore, for $\zeta_1,\zeta_2\in \res{-\Delta}$,
$\zeta_1\ne\zeta_2$, the difference
\[
G(x,y;\zeta_1)-G(x,y;\zeta_2) \sim \frac{\zeta_2-\zeta_1}{8\pi^2}\,
\log|x-y|
\]
is a discontinuous function, so that the singularity cannot be chosen
independent of the spectral parameter.
\end{example}

Therefore, the only non-trivial cases remain $\nu=2$ and $\nu=3$,
which we will consider in the present article.

\begin{example}[\textbf{On-diagonal singularity for the Laplace operator}]
    \label{ex-lb23}
Here we consider the case $A=0$ and $U=0$, i.e. the case of the
Laplace-Beltrami operator $-\Delta$ on the manifold $X$ with
$\nu=2$ or $\nu=3$. Denote the Green function of $-\Delta$ by
$G(x,y;\zeta)$. Take $y\in X$ and introduce polar coordinates
$(r_y,\omega)$, $r_y=d(x,y)$, $\omega\in \SSS^{\nu-1}$, centered
at $y$, then we have in a normal neighborhood $W_y$ of $y$:
\[
-\Delta \psi =-\frac{\partial^2 \psi}{\partial r_y^2}+
\Big(\frac{\nu-1}{r_y}+\theta_y^{-1}\frac{\partial\theta_y}{\partial
r_y}\Big)\,\frac{\partial \psi}{\partial r_y}\,,
\]
where the function $\theta_y=\theta_y(r_y,\omega)$ is defined in
such a way that in $W_y$, we have
$dx=r_y^{\nu-1}\theta_y(r_y,\omega)dr_y\,d\omega$. Since
$r_y^{\nu-1}\theta(r_y,\omega)$ is the Jacobian for the inverse to
the exponential map in $W_y$, there holds $\theta_y(0,\omega)\ge
c_y>0$ and $\displaystyle \frac{\partial}{\partial
r}\theta_y(0,\omega)=0$ for all $\omega\in \SSS^{\nu-1}$.
Moreover, $\inf\,c_y>0$ as $y$ runs over a compact set in $X$.

Denote now
\[
S(x,y)=
\begin{cases}
\dfrac{1}{2\pi}\,\log\dfrac{1}{d(x,y)},& \quad \nu=2,\\[\bigskipamount]
\dfrac{1}{4\pi\,\, d(x,y)},& \quad \nu=3,\\
\end{cases}
\]
and for a fixed $\zeta\in\res{-\Delta}$ denote
$K(x,y):=G(x,y;\zeta)-S(x,y)$. Then there holds
\begin{equation}
            \label{eq-f-g}
(-\Delta-\zeta)K(\cdot,y)=
\theta^{-1}_y\frac{\partial\theta_y}{\partial
r_y}\frac{\partial}{\partial r_y}S(\cdot,y)-\zeta
S(\cdot,y)=:L(x,y).
\end{equation}
It is clear that $L(\cdot,y)\in L^2(W_y)$,
hence due to the Sobolev embedding theorem, $x\mapsto K(x,y)$ is
continuous in $W_y$. Let us show that really $K(x,y)$ is
continuous in $(x,y)$. To do this, we fix $y_0\in X$ and take
$r_0>0$ such that $B(y_0,2r_0)\subset W_{y_0}$. We prove the
following assertion:

\begin{itemize}
\item[(CM)] {\it the map $B(y_0,r_0)\ni y\mapsto L(\cdot,y)\in
L^2(B(y_0,r_0))$ is continuous with respect to the norm topology
of the space $L^2(B(y_0,r_0))$}.
\end{itemize}

Let $\chi\in C^\infty(X)$ such that ${\rm supp}\,\chi\subset
B(y_0,2r_0)$, $\chi(x)=1$ for $x\in B(y_0,r_0)$, and $0\le
\chi(x)\le 1$ for all $x\in X$. Note that $B(y_0,2r_0)$ is a
normal neighborhood of $y$ for all $y\in B(y_0,2r_0)$, therefore
we can assume that $L(x,y)$ is defined for all $x\in X$ and $y\in
B(y_0,2r_0)$. Extend $L$ by zero for $y\notin B(y_0,2r_0)$ and set
$T(x,y)=\chi(x)\chi(y)L(x,y)$. It is clear that $T\in
\mathcal{K}_{\rm cont}(\alpha,p)$ where $p$ is arbitrary number
with $1\le p\le\infty$, and $\alpha=1$ for $\nu=3$, $\alpha$ is
any strictly positive number for $\nu=2$. Using items (A) and (B)
of Lemma~\ref{main} we can easily show that for every $f\in
L^2(X)$ the mapping $\displaystyle B(y_0,r_0)\ni y\rightarrow
\int_{B(y_0,r_0)}\,L(x,y)f(y)\,dy$ is continuous and the
mapping $\displaystyle B(y_0,r_0)\ni y\rightarrow
\int_{B(y_0,r_0)}\,|L(x,y)|^2\,dy$ is also continuous. This
proves the assertion (CM). Returning to Eq.~(\ref{eq-f-g}) we see
that $K(\cdot,y)$ tends to $K(\cdot,y_0)$ with respect to the
topology of $W^2_2\big(B(y_0,r_0)\big)$. Due to the Sobolev
embedding theorem, this implies a uniform convergence in the ball
$B(y_0,r)$, i.e.,
\[
\lim_{y\to y_0}\sup_{x\in B(y_0,r_0)}\big|K(x,y)-K(x,y_0)\big|=0.
\]
This together with the continuity in $x$ proves the required joint
continuity in $(x,y)$. Therefore, the functions $S(x,y)$ are
suitable on-diagonal singularities of the Laplace operator.

Note that the proof of the separate continuity of the function
$K(x,y)$ is considerably simpler and can be found, e.g., in
\cite{CdV}.

\end{example}

\section{On-diagonal behavior for singular scalar potentials}

Below we will use the notation ${L}^{p+}_\text{loc}(X)=\bigcup_{q>p} {L}^q_\text{loc}(X)$.

\begin{lemma}[\textbf{Singularity is independent of the spectral parameter}]
\label{lem-z}
Let $\nu=2$ or $3$,
$A\in [{C}^\infty(X)]^\nu$, $U\in \mathcal{P}(X)$, $\zeta_1,\zeta_2\in\res{H_{A,U}}$,
then the difference $G_{A,U}(x,y;\zeta_1)-G_{A,U}(x,y;\zeta_2)$ is continuous
in $X\times X$.
\end{lemma}

\begin{proof}
The proof follows from the Hilbert resolvent
identity for the kernels: $R_{A,U}(\zeta_1)-R_{A,U}(\zeta_2)=(\zeta_1-\zeta_2)
R_{A,U}(\zeta_1)R_{A,U}(\zeta_2)$.
The integral kernel $\dint_X G_{A,U}(x,z;\zeta_1)G_{A,U}(z,y;\zeta_2)dz$
of $R_{A,U}(\zeta_1)R_{A,U}(\zeta_2)$
is continuous due to Lemma~\ref{main}(B).
\end{proof}

The previous lemma shows that for fixed $A$ and $U$, the
on-diagonal singularity in question exists; for example, as a
singularity one can take $G_{A,U}(x,y;\zeta_0)$ for a fixed
$\zeta_0\in\res{H_{A,U}}$. Our aim is to understand how the
singularity depends on $A$ and $U$.

The following lemma shows that Green functions of Schr\"odinger operators with smooth potentials
have the same on-diagonal singularity.
\begin{lemma}[\textbf{Singularity for operator with smooth potentials}]
\label{lem-c} Let $\nu=2$ or $3$, $A\in [{C}^\infty(X)]^\nu$,
$U,V\in\mathcal{P}(X)\cap {C}^\infty(\Omega)$, where $\Omega$ is a
domain in $X$, then the difference
$G_{A,U}(x,y;\zeta)-G_{A,V}(x,y;\zeta)$ has a continuous extension
to all points $(x,x)$, $x\in \Omega$. In particular, if
$\Omega=X$, then $G_{A,U}(x,y;\zeta)-G_{A,V}(x,y;\zeta)\in
\KK_\mathrm{cont}(p)$ with arbitrary $p\ge 1$.
\end{lemma}

\begin{proof}
Fix a real $E$ sufficiently close to $-\infty$ and take $x_0\in
\Omega$. We show that in a neighborhood of $(x_0,x_0)$ in $X
\times X$, the difference
$F(x,y;E)=G_{A,U}(x,y\,;E)-G_{A,V}(x,y\,;E)$ is the restriction of
a continuous function in this neighborhood. Due to
Lemma~\ref{lem-z} the same will hold for all values of the
spectral parameter.

Let $\Omega_0$ be a bounded subdomain of $\Omega$ and contain
$x_0$; denote $W=U+\chi_{\Omega_0}(V-U)$; it is clear that $W\in
\mathcal{P}(X)$. Since $W-U$ is bounded with compact support,
one has $R_{A,U}(\zeta)-R_{A,W}(\zeta)=R_{A,U}(\zeta)(W-U)R_{A,W}(\zeta)$,
so that
the difference
\[
G_{A,U}(x,y;E)-G_{A,W}(x,y;E)=
\int_X G_{A,U}(x,z;E) \big(W(z)-U(z)\big)G_{A,W}(z,y;E)dz
\]
 is continuous in
$X\times X$ according to Lemma~\ref{main}(B). It remains to show
that the function $L(x,y)=G_{A,V}(x,y;E)-G_{A,W}(x,y;E)$ is
continuous on $\Omega_0\times\Omega_0$. To do this, let us note
that in the sense of distributions the following equality holds:
\begin{multline}
                        \label{ell}
\big((H_{A,V})_x-E +\overline{(H_{A,V})}_y-E\big)L(x,y)\\
{}= \big(W(x)-V(x)\big)G_{A,W}(x,y;E)
+\big(W(y)-V(y)\big)G_{A,W}(x,y;E)\,,
\end{multline}
where $(H_{A,V})_x$ (respectively, $(H_{A,V})_y$) means that
$H_{A,V}$ acts on the first (respectively, the second) argument in
$L$; the bar means that we change the coefficients in $H_{A,V}$ by the
complex conjugate ones. The operator in the left-hand side of
(\ref{ell}) is elliptic in $\Omega_0\times\Omega_0$ with smooth
coefficients, while the right-hand term vanishes in
$\Omega_0\times\Omega_0$. According to the elliptic regularity
theorem $L$ is continuous in $\Omega_0\times\Omega_0$.
\end{proof}

The following Proposition contains our main result on the dependence
of the on-diagonal singularity on singularities of the scalar potential.

\begin{prop}[\textbf{Preserving the on-diagonal singularity under singular perturbations}]
\label{lem-l3}
Let $\nu=2 $ or $3$, $A\in [{C}^\infty(X)]^\nu$, and $U_1, U_2\in \mathcal{P}(X)$.
If $\nu=3$, assume additionally that $U_1-U_2\in {L}^{3+}_\mathrm{loc}(X)$.
Then the difference $G_{A,U_1}(x,y;\zeta)-G_{A,U_2}(x,y;\zeta)$ is
continuous in $X\times X$ for any $\zeta\in\res{H_{A,U_1}}\cap\res{H_{A,U_2}}$.
\end{prop}

\begin{proof} For the sake of brevity we fix $A$ and remove it
from the notation, i.e. instead of $G_{A,U}$ we will write $G_U$ etc.

First of all, using Lemma~\ref{lmain} we choose functions
$V_1,V_2\in {C}^\infty(X)$ semibounded below such that
$W_j:=U_j-V_j= \sum_{s=1}^{n_j} W_{j,s}$, where $W_{j,s}\in
{L}^{p_{j,s}}$ with $2\le p_{j,s}<\infty$, $s=1,\dots,n_j$,
$j=1,2$.

For $\zeta\in\res{H_{U_1}}\cap\res{H_{U_2}}$
the sets $\mathcal{D}_j:=(H_{U_j}-\zeta){C}_0^\infty(X)$ are dense
in ${L}^2(X)$, because ${C}_0^\infty(X)$ is an essential domain of both
$H_{U_1}$ and $H_{U_2}$. As $\psi\in \mathcal{D}_j$, one has
\begin{equation}
          \label{eq-ls}
R_{U_j}(\zeta)\psi-R_{V_j}(\zeta)\psi=R_{V_j}(\zeta)W_j R_{U_j}(\zeta)\psi.
\end{equation}
As the operators on the both sides of~\eqref{eq-ls} are bounded
and coincide on a dense subset, they coincide everywhere,
i.e.~\eqref{eq-ls} holds for any $\psi\in L^2(X)$. Combining
Lemma~\ref{lem-sws} and Lemma~\ref{main}(B) we conclude that in
the dimension two, the operator on the right-hand side
of~\eqref{eq-ls} has a continuous integral kernel, which together
with Lemma~\ref{lem-c} implies the conclusion of the proposition.

Let us consider the dimension three more carefully. To be shorter,
we remove the dependence of the resolvents on $\zeta$ from the
notation. We have the following chain of equalities:
\begin{align*}
R_{U_1}-R_{U_2}&
=R_{V_1}-R_{V_2}+R_{V_1} W_1
R_{U_1}-R_{V_2} W_2 R_{U_2}\\
&=R_{V_1}-R_{V_2}+R_{V_1} W_1 R_{U_1}
+ R_{V_2} W_2 (R_{U_1}-R_{U_2})
- R_{V_2} W_2 R_{U_1}  \\
&=R_{V_1}-R_{V_2}+
R_{V_2} W_2 (R_{U_1} - R_{U_2})
{}+ R_{V_2} (W_1-W_2) R_{U_1} + (
R_{V_1}-R_{V_2}) W_1 R_{U_1}.
\end{align*}
Therefore, $(1-R_{V_2} W_2)(R_{U_1}-R_{U_2})=:L=A+B+C$, where
$A:=R_{V_1}-R_{V_2}$, $B:=R_{V_2} (W_1-W_2) R_{U_1}$,
$C:=(R_{V_1}-R_{V_2}) W_1 R_{U_1}$.

Due to Lemma~\ref{lem-c}, the operator $A$ has an integral
kernel from $\KK_\mathrm{cont}(p)$ with arbitrary $p$, $p\ge 1$.
Since $W_1-W_2\in{L}^{3+}_\mathrm{loc}(X)$, the operator
$B$ has an integral kernel from $\KK_\mathrm{cont}(\infty)$
due to Theorem~\ref{th-bgp} and the items (A), (B) of
Lemma~\ref{main}. As $R_{V_2}-R_{V_1}\in
\KK_\mathrm{cont}(p)$ with arbitrary $p\ge 1$ (Lemma~\ref{lem-c}),
the integral kernel for $C$ is from
$\KK_\text{cont}(\infty)$ due to Theorem~\ref{th-bgp} again and
the items (A), (C) of~Lemma~\ref{main}. Therefore, the operator
$L$ has an integral kernel
$L(x,y)=L(x,y;\zeta)\in\KK_\mathrm{cont}(\infty)$. Now we note that the
multiplication by $W_{2,s}$ is a continuous mapping from
${L}^\infty(X)$ to ${L}^{p_{2,s}}(X)$. At the same time, as
$G_{V_2}\in\KK_\mathrm{cont}(1,p)$, $p\ge 1$, the resolvent
$R_{V_2}$ is a bounded operator from each
${L}^{p_{2,s}}(X)$ to ${L}^\infty(X)$ due to Lemma~\ref{lem-ka}.
Since $L=(1-R_{V_2} W_2 ) (R_{U_1}-R_{U_2})$, we can combine
Theorem~\ref{th-bgp} and Lemma~\ref{lem-ka} to show that the
operator $L$ is a bounded map from ${L}^{p}(X)$ to
${L}^\infty(X)$ for any $p$ with $3/2<p<\infty$. Since
$\big|L(x,y;\zeta)\big|=\big|L(y,x;\bar \zeta)\big|$, we see from \eqref{est-s}
that $L(x,y)\in\KK_\mathrm{cont}(q)$ for any $q$ with
$1<q<3$.

One can find $\zeta$ such that $\|R_{V_2}(\zeta)
W_2\|_{\infty,\infty}=:\alpha<1$ (see \cite{BGP}),
therefore, the operator $1-R_{V_2} W_2$ acting in
${L}^\infty(X)$ is invertible and for any $n\in\NN$ there holds
\begin{equation}
        \label{gu1u2}
R_{U_1}-R_{U_2}=\sum_{k=0}^{n-1} (R_{V_2} W_2)^k L
+ (1-R_{V_2} W_2)^{-1} (R_{V_2} W_2)^n L.
\end{equation}
Applying iteratively Lemmas~\ref{lem-sws} and~\ref{main}(A) and
taking into account Theorem~\ref{th-bgp}, we can show that the
operators $(R_{V_2} W_2)^k R_{V_2}$ have integral
kernels from $\KK_\mathrm{cont}(\beta_k,\infty)$ with $\beta_k\le
1$. At the same time, all these operators are bounded from
${L}^p(X)$ to ${L}^\infty(X)$ for any $p$ with $3/2<p<\infty$.
Using the same arguments as for $L$ above, we conclude that
these kernels are in $\KK_\mathrm{cont}(\beta_k,q)$ for any $q$
with  $1<q<3$. Applying now Lemma~\ref{main}~(C) one proves that
the first term on the right-hand side has a continuous integral
kernel.

Denote $T_n:=(1-R_{V_2} W_2)^{-1} (R_{V_2} W_2)^{n-1} R_{V_2}$;
this operator is bounded from each ${L}^{p_{j,s}}(X)$ to
${L}^\infty(X)$; due to the Gelfand-Dunford-Pettis theorem, this
is an integral operator with an integral kernel $T_n(x,y)$. The
second term in~\eqref{gu1u2} takes the form $T_n W_2 L$, and by
virtue of Lemma~\ref{lem-sws} this is also an integral operator
with the kernel $S_n(x,y):=\dint_X T_n(x,z) W_2(z) L(z,y)\,dz$.
{}From the other side, one can write $S_n(x,y)=T_n  W_2 l_y(x)$,
where $l_y(x):=L(x,y)$. Note that for each $y\in X$ there holds
$l_y\in{L}^\infty(X)$, and the operator $T_n W_2$ is a bounded
mapping from ${L}^\infty(X)$ to ${L}^\infty(X)$ with the norm
$\|T_n W_2\|_{\infty,\infty}\le \big\|(1-R_{V_2}
W_2)^{-1}\big\|_{\infty,\infty}\cdot \|R_{V_2}
W_2\|^n_{\infty,\infty}\le\alpha^n/(1-\alpha)$.

Now let us fix $x_0\in X$ and take a bounded open neighborhood
$\Omega$ of $x_0$. It is clear that $\|l_y\|_\infty\le c_\Omega$
for all $y\in\Omega$ with a certain $c_\Omega>0$. Therefore
$\sup_{x,y\in\Omega}|S_n(x,y;\zeta)|\le
c_\Omega\alpha^n/(1-\alpha)$. Take $\epsilon>0$ and choose $n$
such that $c_\Omega\alpha^n/(1-\alpha)<\epsilon$. From
Eq.~\eqref{gu1u2} we have in $\Omega\times\Omega$ the relation
$G_{U_1}(x,y;\zeta)-G_{U_2}(x,y;\zeta)= K_n(x,y) +\tau_n(x,y)$,
where $K_n$ is continuous and $|S_n|<\epsilon$. As $\epsilon$ is
arbitrary, this means that $G_{U_1}(x,y;\zeta)-G_{U_2}(x,y;\zeta)$
is continuous in $\Omega\times\Omega$. Since $x_0\in X$ is
arbitrary, the lemma is proven. Due to Lemma~\ref{lem-z}, this
holds for all $\zeta\in\res{H_{V_1}}\cap\res{H_{V_2}}$.
\end{proof}

The following example shows that the condition
$U_1-U_2\in{L}^{3+}_\mathrm{loc}(X)$ can not be omitted
in dimension three.
\begin{example}[\textbf{Coulomb potential in three dimensions}]
 \label{ex-c}
Let $X=\RR^3$, $A=0$, and $U=q/|x|$, i.e.
$H\equiv H_{A,U}=-\Delta+q/|x|$.
Clearly, $U\notin {L}^{3+}_\mathrm{loc}(\RR^3)$. The Green function
can be calculated explicitly~\cite{GS}:
\begin{equation}
                     \label{Coul6}
G(x,y;\zeta)=\frac{\Gamma(1-\kappa)}{4\pi|x-y|}
\bigg[W_{\kappa,1/2}(\sqrt{-\zeta}\,\xi)M'_{\kappa,1/2}(\sqrt{-\zeta}\,\eta)
- W'_{\kappa,1/2}(\sqrt{-\zeta}\,\xi)M_{\kappa,1/2}(\sqrt{-\zeta}\,\eta)\bigg],
\end{equation}
where  $\xi:=|x|+|y|+|x-y|$,
$\eta:=|x|+|y|-|x-y|$, $\kappa=-q/\sqrt{-4\zeta}$,
$M_{\kappa,1/2}$ and $W_{\kappa,1/2}$ are the
Whittaker functions,
\begin{equation}
                     \label{Coul8}
M_{\kappa,1/2}(x)=e^{x/2}x\Phi(a,2;x)\,,\quad\,W_{\kappa,1/2}(x)=e^{x/2}x\Psi(a,2;x)\,.
\end{equation}
Here $\Phi(a,c;x)$ and $\Psi(a,c;x)$ are the Kummer function and
the Tricomi function, respectively. We prove in
Appendix~\ref{ap-a} the asymptotics
\begin{multline} \label{green-coulomb}
G(x,0;\zeta)=\frac{1}{4\pi |x|}+\frac{q}{4\pi}\log |x|
-\frac{\sqrt{-\zeta}}{4\pi}\\
{}+
\frac{q}{4\pi}\left(\psi\left(1+\frac{q}{2\sqrt{-\zeta}}\right)+\log\sqrt{-\zeta}+\log(2/e)+2C_E\right)
+O(|x|\log |x|).
\end{multline}
Therefore, the singularity for $G(x,y;\zeta)$ contains an unavoidable logarithmic term
and is different from the standard three-dimensional singularity.
\end{example}

\section{Dependence of the singularity on the magnetic field}

\begin{lemma}[\textbf{Singularity due to the magnetic field in two dimensions}]
    \label{pr2d}
Let $\nu=2$, then for any $A\in[{C}^\infty(X)]^\nu$ the difference
$G_{A,0}(x,y;\zeta)-G_{0,0}(x,y;\zeta)$ is continuous in $X\times
X$ if $\zeta\in\res{H_{A,0}}\cap\res{H_{0,0}}$.
\end{lemma}

\begin{proof}
Let $x_0$ be an arbitrary point of $X$. We show that the
difference $G_{A,0}(x,y;\zeta)-G_{0,0}(x,y;\zeta)$ is continuous
in a neighborhood of $(x_0,x_0)$ for at least one value of the
spectral parameter $\zeta$; due to Lemma~\ref{lem-z} this
difference is continuous for all admissible spectral parameters.

Take two sufficiently small numbers $r$ and $r_0$ with $0<r<r_0$.
Fix a function $\phi\in C_0^\infty(X)$ such that $\supp
\phi\subset B(x_0,r_0)$, $\phi(x)=1$ as $x\in B(x_0,r)$. Denote
for brevity $H_0:=H_{0,0}$, $H_1:=H_{A,0}$, $H_2:=H_{\phi A,0}$;
the corresponding Green functions will be denoted by $G_0$, $G_1$,
and $G_2$, respectively.

In $B(x_0,r)\times B(x_0,r)$ for real $\zeta$ sufficiently close
to $-\infty$ one has in the sense of distributions
\[
\Big(\big((H_1)_x-\zeta\big)+\big((\overline{H_2})_y-\zeta\big)\Big)
\Big(G_1(x,y;\zeta)-G_2(x,y;\zeta)\Big)=0,
\]
therefore, due to the elliptic regularity, the difference
$G_1(x,y;\zeta)-G_2(x,y;\zeta)$ is continuous in  $B(x_0,r)\times
B(x_0,r)$. Now we are going to show that
$G_2(x,y;\zeta)-G_0(x,y;\zeta)$ is continuous. Since $H_0$ and
$H_2$ are uniformly elliptic operators with $C^\infty$-bounded
coefficients, we are able to use estimates for the Green functions
and their derivatives obtained in~\cite{Shub}. First of all,
\begin{equation}
           \label{green-k-2}
G_0(x,y;\zeta),\, G_2(x,y;\zeta)\in \KK_\text{cont}(\lambda,q)
\end{equation}
for arbitrary $\lambda>0$ and $q\in[1,\infty]$ (see
Theorem~\ref{th-bgp}). Moreover, for $\zeta$ close to $-\infty$
both these kernels are smooth outside the diagonal $x=y$, and
according to ~\cite[Theorem A1.3.7]{Shub} we have
\[
\bigg|\partial_x G_0(x,y;\zeta)\bigg| \le C
\bigg(1+\frac{\big|\log d(x,y)\big|}{d(x,y)}\bigg) e^{-\omega
d(x,y)}, \quad j=1,2\,,
\]
where $\partial$ is any first order derivative
taken in canonical coordinates,
and $C,\omega>0$. Additionally, by~\cite[Theorem A1.2.3]{Shub}
for any $p\ge 1$ there exist $\epsilon,\,C'>0$ such that
\[
\sup_x \int_{d(x,y)>r} \bigg|\partial_x G_0(x,y;\zeta)\bigg|^p
e^{\epsilon d(x,y)}dy +\sup_y \int_{d(x,y)>r} \bigg|\partial_x
G_0(x,y;\zeta)\bigg|^p e^{\epsilon d(x,y)}dx \le C', \quad
j=1,2\,.
\]
This implies the inclusion
\begin{equation}
  \label{green-k}
\partial_x  G_0(x,y;\zeta)\in\KK_\text{cont}(1+\lambda,q),\quad
\end{equation}
with the same $\lambda$ and $q$ as in~\eqref{green-k-2}.

In canonical coordinates in $B(x_0,r_0)$ both $H_0$ and $H_2$ are
given by symmetric second-order elliptic expressions with the same
principal symbol, in particular, the difference $T:=H_2-H_0$ is
defined by a first order differential expression,
$T=b_1(x)\,\partial_1+b_2(x)\, \partial_2 +c(x)$, where $b_1$,
$b_1$, $c$ are compactly supported smooth functions. For the
functions of the form $\psi=(H_0-\zeta)\phi$ with
$\phi\in{C}_0^\infty(X)$ we have $(H_2-\zeta) \phi=(H_0+T-\zeta)
R_0(\zeta)\psi=\big(1+T R_0(\zeta)\big)\psi$, therefore,
$R_0(\zeta)\psi-R_2(\zeta)\psi=R_2(\zeta)T R_0(\zeta)\psi$. In
terms of integral kernels this means
\begin{multline}
         \label{dif-g}
\int_X G_0(x,y;\zeta)\psi(y)dy-\int_X G_2(x,y;\zeta)\psi(y)dy\\
{}=\int_X G_2(x,z;\zeta)\bigg[
b_1(z)\,\partial_1+b_2(z)\,\partial_2
+c(z)\bigg]\int_X G_0(z,y;\zeta) \psi(y)\,dy\,dz\\
{}=\int_X G_2(x,z;\zeta)\int_X \bigg[ b_1(z)K_1(z,y;\zeta)
+b_2(z) K_2(z,y;\zeta){}
+c(z)G_0(z,y;\zeta)\bigg]\psi(y)\,dy\,dz,
\end{multline}
where
\[
K_1(z,y;\zeta):=\partial_{z_1} G_0(z,y;\zeta), \quad
K_2(z,y;\zeta):=\partial_{z_2} G_0(z,y;\zeta).
\]
According to the general theory of elliptic operators, the set
$(H_0-\zeta){C}_0^\infty(X)$ is dense in all ${L}^p(X)$ with any
$p$, $1\le p<\infty$, if $\zeta$ is sufficiently close to
$-\infty$~\cite[Section~A1.2]{Shub}. Due to the
estimates~\eqref{green-k-2}, \eqref{green-k}, and
Lemma~\ref{lem-ka}, the kernels $K_1$ and $K_2$ define bounded
operators from ${L}^q(X)$ to ${L}^\infty(X)$ for arbitrary $q>2$;
denote these operators by $K_1(\zeta)$ and $K_2(\zeta)$. In this
notation, the expression in the right-hand side of~\eqref{dif-g}
can be rewritten as
\[
R_0(\zeta)\psi-R_2(\zeta)\psi {}= \big[R_2(\zeta) b_1
K_1(\zeta)+R_2(\zeta) b_2 K_2(\zeta) +R_2(\zeta)\, c\,
R(\zeta)\big]\psi.
\]
The operators in the both sides are bounded from ${L}^q(X)$ to ${L}^\infty(X)$
with any $q>2$ and coincide on a dense subset, therefore, the corresponding
kernels coincide, i.~e.
\begin{multline}
      \label{eq-g-gb}
G_0(x,y;\zeta)-G_2(x,y;\zeta)=
\int_X G_2(x,z;\zeta)b_1(z)K_1(z,y;\zeta)dz\\
{}+\int_X G_2(x,z;\zeta)b_2(z)K_2(z,y;\zeta)dz+ \int_X
G_2(x,z;\zeta)c(z)G_0(z,y;\zeta)dz\,.
\end{multline}
By Lemma~\ref{main}~(B), the function on the right-hand side of
(\ref{eq-g-gb}) is continuous.
\end{proof}

The three-dimensional analogue of Lemma~\ref{pr2d} is not true as
the following example shows.


\begin{example}[\textbf{Three-dimensional Landau Hamiltonian}]
   \label{ex-3d}
Consider in ${L}^2(\RR^3)$ the vector potential of a non-zero
uniform magnetic field. By a suitable choice of coordinates one
can assume that the field is directed along the $x_3$-axis, i.e.
the magnetic strength vector is $\mathbf{B}=(0,0,2\pi \xi x_3)$,
where $\xi>0$ is the density of the magnetic flux through the
plane $(x_1,x_2)$. Choose the symmetric gauge for the the magnetic
vector potential, $\mathbf{A}(\mathbf{x})=\frac{1}{2}
\mathbf{B}\times \mathbf{x}$, then $H:=H_{\mathbf{A},0}$ takes the
form
\[
H=\Big(i\frac{\partial}{\partial x_1}-\pi\xi x_2\Big)^2+
\Big(i\frac{\partial}{\partial x_2}+\pi\xi x_1\Big)^2
-\frac{\partial^2}{\partial x_3^2},
\]
and the corresponding Green function is
$G(\mathbf{x},\mathbf{y};\zeta)=\Phi(\mathbf{x},\mathbf{y})F(\mathbf{x}-\mathbf{y};\zeta)$,
where
\begin{equation}
F(\mathbf{x};\zeta)=\int_0^\infty\frac{\exp[-\pi|\xi|(\mathbf{x}_\bot^2(e^t-1)^{-1}+
\mathbf{x}_{\|}^2t^{-1}]}{(1-e^{-t})\exp\bigg[\Big(\dfrac{1}{2}-
\dfrac{\zeta}{4\pi|\xi|}\Big)t\bigg]\sqrt{t}}\,dt\,,
                 \label{UF9}
\end{equation}
$\mathbf{x}_\bot=(x_1,x_2,0)$ and $\mathbf{x}_{\|}=(0,0,x_3)$
\cite{GD96}. In Appendix~\ref{app-b} we prove the asymptotics
\begin{equation}
                           \label{T3.5}
G(\mathbf{x},\mathbf{y};\zeta)=\frac{e^{i
\pi\xi(\mathbf{x}_\bot\wedge
\mathbf{y}_\bot)}}{4\pi|\mathbf{x}-\mathbf{y}|}+
\frac{1}{4}\left(\frac{|\xi|}{\pi}\right)^{1/2} {\rm
Z}\left(\frac{1}{2}\,;\,
\frac{1}{2}-\frac{\zeta}{4\pi|\xi|}\right)+o(|\mathbf{x}-\mathbf{y}|)
\end{equation}
as $|\mathbf{x}-\mathbf{y}|\to 0$; here ${\rm Z}(z;u)$ is the
generalized Riemann $\zeta$-function (also known as the Hurwitz
$\zeta$-function). Therefore, the on-diagonal asymptotics is
\[
S(\mathbf{x},\mathbf{y})=\frac{e^{i \pi\xi(\mathbf{x}_\bot\wedge
\mathbf{y}_\bot)}}{4\pi|\mathbf{x}-\mathbf{y}|}=
\frac{1}{4\pi|x-y|}\,\exp\Big(\frac{i\mathbf{B}(\mathbf{x}\times\mathbf{y})}{2}\Big)\,.
\]
\end{example}

\section{Summary of results}

We summarize some corollaries from the proven assertions in the
following two theorems.

\begin{theorem}[\textbf{On-diagonal singularities of the Green functions in dimension two}]
\label{main-th-2d}
 On a two-dimensional manifold of bounded geometry $X$,
for any vector potential $A\in [{C}^\infty(X)]^2$ and
scalar potential $U\in \mathcal{P}(X)$, the Green function $G_{A,U}$
of the Schr\"odinger operator $H_{A,U}=-\Delta_A+U$ has the same on-diagonal
singularity as that for the Laplace-Beltrami operator, i.e.
\[
G_{A,U}(x,y;\zeta)=\dfrac{1}{2\pi}\log \frac{1}{d(x,y)}+
G^\mathrm{ren}_{A,U}(x,y;\zeta),
\]
where $G^\mathrm{ren}_{A,U}$ is continuous on $X\times X$.
\end{theorem}

\begin{proof}
Proposition~\ref{lem-l3} shows that the singularity does not depend
on the scalar potential $U\in\mathcal{P}(X)$, and Lemma~\ref{pr2d} shows that
it is independent of the magnetic potential.
Therefore, the singularity coincides with that for the Laplacian, see Example~\ref{ex-lb23}.
\end{proof}

\begin{theorem}[\textbf{On-diagonal singularities of the Green functions in dimension three}]
  \label{main-th-3d}

Let $X$ be a three-dimensional manifold of bounded geometry. For
$U\in\mathcal{P}(X)$ and $A\in[C^\infty(X)]^3$ consider the
Schr\"odinger operator $H_{A,U}=-\Delta_A+U$ and its Green
function $G_{A,U}(x,y;\zeta)$. If $U_1,U_2\in\mathcal{P}(X)$ and
$U_1-U_2\in L^{3+}_\mathrm{loc}(X)$, then the Green functions
$G_{A,U_1}$ and $G_{A,U_2}$ have the same on-diagonal singularity
$($i.e. $G_{A,U_1}-G_{A,U_2}$ is continuous in $X\times X$$)$. In
particular, for any $U\in\mathcal{P}(X)\cap
L^{3+}_\mathrm{loc}(X)$ there holds
\begin{equation}
 \label{eq-lb3}
G_{0,U}(x,y;\zeta)=\frac{1}{4\pi\,d(x,y)}+G^\mathrm{ren}_{0,U}(x,y;\zeta),
\end{equation}
where $G^\mathrm{ren}_{0,U}$ is continuous in $X\times X$.
\end{theorem}

\begin{proof} The theorem is a simple corollary of Proposition~\ref{lem-l3}, and
the formula~\eqref{eq-lb3} follows from Example~\ref{ex-lb23}.
\end{proof}

\begin{rem} Contrary to the two-dimensional case, the singular
term of the Green function for the three-dimensional Schr\"odinger
operator $H_{A,U}$ \emph{does} depend on the scalar potential $U$
as well as on the magnetic vector potential $A$. In particular, if
$A$ is the vector potential of a uniform magnetic field
$\mathbf{B}$ in $X=\RR^3$, then instead of \eqref{eq-lb3} we have
\[
G_{A,0}(\mathbf{x},\mathbf{y};\zeta)=
\frac{1}{4\pi|x-y|}\,\exp\Big(\frac{i\mathbf{B}(\mathbf{x}\times\mathbf{y})}{2}\Big)
+G^\mathrm{ren}_{A,0}(\mathbf{x},\mathbf{y};\zeta)\,,
\]
see Example~\ref{ex-3d}. On the other hand, the dependence on
scalar potentials is shown in Example~\ref{ex-c}.
\end{rem}

\section*{Acknowledgments}
The authors thank Gilles Carron, Pavel Exner, Daniel Grieser, and
Evgeny Korotyaev for fruitful discussions. The work was partially
supported by the Deutsche Forschungsgemeinschaft, INTAS, and the
Russian Foundation for Basic Research (Grant no.~02-01-00804).

\appendix

\section{Off-diagonal asymptotics for the inverse harmonic oscillator
in dimension one}\label{app-c}

The Green function $G(x,y;\zeta)$ for the inverse harmonic oscillator
$H=-d^2/dx^2-\omega^2 x^2/4$,
has the form
\begin{multline}
G(x,y;\zeta)=\frac{e^{i\pi/4}\Gamma\left(\frac{1}{2}-i\zeta\right)}{\sqrt{2\pi\omega}}\\
        \label{17}
{}\times U\big(-i\zeta/\omega,\,e^{-i\pi/4}\omega^{1/2}\max(x,y)\big)\times
U\big(-i\zeta/\omega,\,e^{-i\pi/4}\omega^{1/2}\max(-x,-y)\big),
\end{multline}
where $\Im\zeta>0$ and $U(a,x)$ is the Weber function, see~\cite[Chapter~19]{AS}.
Using \cite[no.~19.8.1]{AS}, for large $z$ one obtans $U(a,z)=e^{-z^2/4}z^{-\frac12-a}\,u(z)$,
where $\lim_{z\to\infty} u(z)=1$. Returning to the Green function we see that
for fixed $x$ and large $y$ one has (assuming $y>x$)
\[
G(x,y;\zeta)=\frac{e^{i\pi/4}\Gamma(\tfrac12-i\zeta)}{\sqrt{2\pi\omega}}\,
U\big(-i\zeta/\omega,\,-e^{-i\pi/4}\omega^{1/2}\,x\big)
\frac{e^{i\omega y^2/4}}{\big(e^{-i\pi/4}\omega^{1/2}\,y\big)^{-i\zeta/\omega+\frac12}}\,v(y),
\]
where $\lim_{y\to\infty} v(y)\ne 0$.
Therefore, for large $|x-y|$ the Green function
has only a polynomial decaying.

\section{On-diagonal singularity for the Coulomb Hamiltonian}\label{ap-a}
Here we prove the asymptotics~\eqref{green-coulomb}.

We are interested in asymptotics of the functions
$x\mapsto G(x,x_0;\zeta)$ as $x\to x_0$ at fixed
$\zeta\in\res{H}$ and $x_0\in\RR^3$. As the potential is smooth outside the origin,
the Green function has the standard on-diagonal asymptotics
if $x_0\ne 0$. We consider the case $x_0=0$. We have
$M_{\kappa,1/2}(0)=0$, $M'_{\kappa,1/2}(0)=1$, therefore,
\begin{equation*}
G(x,0;\zeta)=\frac{\Gamma(1-\kappa)}{4\pi |x|}W_{\kappa,1/2}(2\sqrt{-\zeta}|x|).
\end{equation*}
Consider the following expansions (cf. items 6.1(1) and 6.8(13) in \cite{BE}):
\begin{align*}
\Phi(a,2;x)&=1+\frac{a}{2}x+\frac{a(a+1)}{12}x^2+\ldots\,,\\
\Psi(a,2;x)&=\frac{1}{x\Gamma(a)}+\Phi(a,2;x)\log x
+\sum\limits_{k=0}^{\infty}
\frac{\Gamma(a+k)[\psi(a+k)-\psi(1+k)-\psi(2+k)]}{\Gamma(a)(k+1)!k!}x^k\notag\\
&=A_{-1}x^{-1}+A_0+A_1x+A_2x^2+\ldots+B_0\log x+ B_1x\log
x+B_2x^2\log x+\ldots\,,
\end{align*}
where
\begin{gather*}
A_{-1}=\frac{1}{\Gamma(a)},\quad
A_0=\frac{\psi(a)-\psi(1)-\psi(2)}{\Gamma(a-1)},\quad
A_1=\frac{a(\psi(a+1)-\psi(2)-\psi(3))}{2\,\Gamma(a-1)},\\
A_2=\frac{a(a+1)(\psi(a+2)-\psi(3)-\psi(4))}{12\,\Gamma(a-1)},\,\,
B_0=\frac{1}{\Gamma(a-1)},\,\,
B_0=\frac{a}{2\,\Gamma(a-1)},\,\,
B_2=\frac{a(a+1)}{12\,\Gamma(a-1)}.
\end{gather*}
Using (\ref{Coul8}), we get
\begin{align*}
W_{\kappa,1/2}(x)&=A_{-1}+\left(A_0-\frac{1}{2}A_{-1}\right)x+B_0x\log
x +O(|x^2\log x|)\\
&=
\frac{1}{\Gamma(a)}+\left(\frac{\psi(a)-\psi(1)-\psi(2)}{\Gamma(a-1)}-\frac{1}{2\,\Gamma(a)}\right)\,x+
\frac{1}{\Gamma(a-1)}x\log x+O(|x^2\log x|)\,.
\end{align*}
Since $\psi(1)=-C_E$, $\psi(2)=1-C_E$, where $C_E$ is
the Euler constant, we get~\eqref{green-coulomb} after some trivial algebra.

\section{On-diagonal singularity of the three-dimensional Landau Hamiltonian}\label{app-b}
In this appendix, we are going to prove the asymptotics~\eqref{T3.5}.

Set in the integral \eqref{UF9} $\mathbf{x}_\bot=0$ and denote
$\mathbf{x}_{\|}=z$. Then after the change of variables $t\to t^2$
in this integral, we obtain
\begin{equation}
               \label{As1}
G(0,0,z;0,0,0;\zeta)=\frac{|\xi|^{1/2}}{2\pi}\,\int_0^\infty\frac{\exp(-az^2t^{-2}
-ct^2)}{1-e^{-t^2}}\,dt\,,
\end{equation}
where $a=\pi|\xi|$ and $\displaystyle
c=\frac{1}{2}-\frac{\zeta}{4\pi|\xi|}$. Represent now
$G(0,0,z;0,0,0;\zeta)=f_1(z;\zeta)+f_2(z;\zeta)$, where
\begin{gather}
f_1(z;\zeta)=\frac{|\xi|^{1/2}}{2\pi}\,\int_0^\infty
\frac{\exp(-az^2t^{-2} -ct^2)}{t^2}\,dt\,,\notag\\
                      \label{comp0}
f_2(z;\zeta)=\frac{|\xi|^{1/2}}{2\pi}\,\int_0^\infty
\left(\frac{1}{1-e^{-t^2}}-\frac{1}{t^2}\right)\,\exp(-az^2t^{-2}
-ct^2)\,dt\,.
\end{gather}
Changing the variable $t\to t^{-1}$ and using the relation
\[
\dint_0^\infty
\exp(-bt^{2}-c/t^2)\,dt\,=\,\frac{1}{2}(\pi/b)^{1/2}\exp(-2(bc)^{1/2})
\]
(see \cite{PBM}, V.~I, Formula~ 2.3.16.3), we obtain
$f_1(z;\zeta)=\exp\big(-(2\pi|\xi|-\zeta)^{1/2}|z|\big)/\big(4\pi|z|\big)$,
or $G(0,0,z;0,0,0;\zeta)=\big(4\pi|z|\big)^{-1}+g(z;\zeta)$,
where
\begin{equation}
                      \label{comp}
g(z;\zeta)=-\frac{1}{4\pi}(2\pi|\xi|-\zeta)^{1/2}+f_2(z;\zeta)\,.
\end{equation}
It is clear that the function $g$ is continuous with respect to
$z$ and analytic with respect to $\zeta$, $\zeta\in\res{H_{A,0}}$.
We can rewrite \eqref{As1} in the form
\begin{equation}
               \label{As6}
\frac{|\xi|^{1/2}}{2\pi}\,\int_0^\infty
\frac{\exp(-\pi|\xi|z^2t^{-1})}{(1-e^{-t})\exp\bigg(\Big(\frac{1}{2}
-\frac{\zeta}{4\pi|\xi|}\Big)t\bigg)\sqrt{t}}\,dt=\frac{1}{4\pi|z|}+g(z;\zeta)\,.
\end{equation}
Let $h(t)=(e^t-1)^{-1}-t^{-1}$; the function $h$ is real-analytic
on the whole line, $h(t)\to 0$ as $t\to+\infty$ and $h(t)\to -1$
as $t\to-\infty$. Therefore, $h$ is bounded on $\RR$. Let us
represent $F(\mathbf{x};\zeta)$ in the form
\begin{equation}
                           \label{As7}
\begin{gathered}
F(\mathbf{x};\zeta)=\int_0^\infty\frac{\exp(-\pi|\xi|\mathbf{x}^2t^{-1})}
{(1-e^{-t})\exp\left[\displaystyle\left(\frac{1}{2}-
\frac{\zeta}{4\pi|\xi|}\right)t\right]\sqrt{t}}\,dt
\\{}+
\int_0^\infty\frac{\exp(-\pi|\xi|\mathbf{x}^2t^{-1})}
{(1-e^{-t})\exp\left[\displaystyle\left(\frac{1}{2}-
\frac{\zeta}{4\pi|\xi|}\right)t\right]\sqrt{t}}\left\{\exp[-\pi|\xi|\mathbf{x}_\bot
h(t)]-1\right\} \,dt
{}\equiv I_1(\mathbf{x},\zeta)+I_2(\mathbf{x},\zeta).
\end{gathered}
\end{equation}
It is easy to show that $I_2$ is a continuous function in the
domain $\mathbf{x}\in\RR^3$, ${\rm Re}\,\zeta<2\pi|\xi|$. Let us
show that $I_2(\mathbf{x},\zeta)\to 0$ locally uniformly with
respect to $\zeta$, ${\rm Re}\,\zeta<2\pi|\xi|$, as $\mathbf{x}\to
0$. It is sufficient to show that
\[
A(\mathbf{x},\zeta)\equiv
\int_0^\infty\frac{\exp(-\pi|\xi|\mathbf{x}^2t^{-1})}
{(1-e^{-t})\exp\left[\displaystyle\left(\frac{1}{2}-
\frac{\zeta}{4\pi|\xi|}\right)t\right]\sqrt{t}}\left|\exp[-\pi|\xi|\mathbf{x}^2_\bot
h(t)]-1\right| \,dt\to 0
\]
locally uniformly with respect to $\zeta\in\RR$,
$\zeta<2\pi|\xi|$, as $\mathbf{x}\to 0$. Fix $\zeta\in\RR$,
$\zeta<2\pi|\xi|$. Since $\mathbf{x}^2_\bot\le \mathbf{x}^2$, we
have $\left|\exp[-\pi|\xi|\mathbf{x}^2_\bot h(t)]-1\right|\le {\rm
const}\,\mathbf{x}^2$ in a neighborhood of the point $(0,0,z)$.
Therefore, using \eqref{As6}, we get
\[
 A(\mathbf{x},\zeta)\le c\mathbf{x}^2
\int_0^\infty\frac{\exp(-\pi|\xi|\mathbf{x}^2t^{-1})}
{(1-e^{-t})\exp\left[\displaystyle\left(\frac{1}{2}-
\frac{\zeta}{4\pi|\xi|}\right)t\right]\sqrt{t}} \,dt \le
\frac{|\mathbf{x}|}{|\xi|^{1/2}}+\frac{c\mathbf{x}^2}{|\xi|^{1/2}}f(|\mathbf{x}|,\zeta)\,,
\]
and we get the required limit. Using \eqref{As6} again, we obtain
\begin{equation}
                           \label{As9}
I_1(|\mathbf{x}|,\zeta)=\frac{1}{|\xi|^{1/2}|\mathbf{x}|}+f(|\mathbf{x}|,\zeta)\,.
\end{equation}
Using \eqref{As7} and \eqref{As9} we get
\[
G(\mathbf{x},\mathbf{y};\zeta)=\frac{1}{4\pi}\frac{\exp\left[\pi
i\xi(\mathbf{x}_\bot\wedge
\mathbf{y}_\bot)\right]}{|\mathbf{x}-\mathbf{y}|}+\tilde
F(\mathbf{x},\mathbf{y};\zeta)\,,
\]
where $\tilde F(\mathbf{x},\mathbf{y};\zeta)$ is jointly
continuous with respect to
$(\mathbf{x},\mathbf{y})\in\RR^3\times\RR^3$ for all
$\zeta\in\res{H_{\mathbf{A},0}}$.

Denote $Q(\zeta)=\lim_{|\mathbf{x}-\mathbf{y}|\to0}\,\tilde
F(\mathbf{x},\mathbf{y};\zeta)$; this limit is independent of
$\mathbf{x}$ and $\mathbf{y}$ since $\tilde
F(\mathbf{x},\mathbf{y};\zeta)$ is invariant with respect to
magnetic translations $T_\mathbf{a}$, $\mathbf{a}\in
\mathbb{R}^3$: $T_\mathbf{a}f(\mathbf{x})=\exp[\pi
i\xi(\mathbf{a}_\bot\wedge
\mathbf{x}_\bot)]\,f(\mathbf{x}-\mathbf{a})$. From (\ref{UF9}) we
obtain
\[
\frac{\partial}{\partial \zeta}Q(\zeta)=
\frac{1}{16\pi^2|\xi|^{1/2}}\int_0^\infty\,\exp\left[\displaystyle\left(\frac{1}{2}-
\frac{\zeta}{4\pi|\xi|}\right)t\right]\,(1-e^{-t})^{-1}\,\sqrt{t}\,dt\,.
\]
Using Equation (1.10.4) from \cite{BE} we get
$\dint_0^\infty\,t^{s-1}\,e^{-vt}\,(1-e^{-t})^{-1}\,dt=\Gamma(s)\,{\rm
Z}(s,v)$ and the obvious relation $\partial{\rm Z}(s,v)/\partial
v=-s\,{\rm Z}(s+1,v)$ implies immediately
\begin{equation}
                      \label{comp1}
Q(\zeta)=\frac{1}{4}\left(\frac{|\xi|}{\pi}\right)^{1/2} {\rm
Z}\left(\frac{1}{2}\,;\,\frac{1}{2}-\frac{\zeta}{4\pi|\xi|}\right)+C
\end{equation}
with a constant $C\in \mathbb{R}$. To determine $C$ we compare
(\ref{comp1}) with (\ref{comp}) in the limit $\Re\zeta\to-\infty$.
Since $Q(\zeta)=g(0;\zeta)$, we have from (\ref{comp}) and
(\ref{comp0}):
\[
Q(\zeta)- \frac{1}{4\pi}(2\pi|\xi|-\zeta)^{1/2}\,  \to 0\quad
\text {as}\quad \Re\zeta\to-\infty\,.
\]
On the other hand, by the Hermite relation (see (1.10.7)
from~\cite{BE}) there holds ${\rm Z}\left(1/2,v\right)+2v^{1/2}\to
0$ as $\Re v\to+\infty$. Comparing the two last relations with
(\ref{comp1}), we get $C=0$. Thus, (\ref{T3.5}) is proven. Note
that the expression for $Q(\zeta)$ was obtained at the physical
level of rigor in \cite{DD} and can be found also in \cite{DO}.

\end{document}